\documentclass[prb,twocolumn,amsmath,amssymb,superscriptaddress]{revtex4}
\usepackage[pdftex]{graphicx}
\usepackage{mathrsfs}
\usepackage[sort&compress]{natbib}
\newcommand{\DD}{\overline{D}}
\begin{document}
\title{Electronic compressibility of layer polarized bilayer graphene}
\author{A.~F.~Young}
    \affiliation{Department of Physics, Columbia University, New York, New York 10027, USA}
\author{C.~R.~Dean}
    \affiliation{Department of Electrical Engineering, Columbia University, New York, New York 10027, USA}
    \affiliation{Department of Mechanical Engineering, Columbia University, New York, New York 10027, USA}
\author{I.~Meric}
    \affiliation{Department of Electrical Engineering, Columbia University, New York, New York 10027, USA}
\author{S.~Sorgenfrei}
    \affiliation{Department of Electrical Engineering, Columbia University, New York, New York 10027, USA}
\author{H.~Ren}
    \affiliation{Department of Physics, Columbia University, New York, New York 10027, USA}
\author{K.~Watanabe}
\author{T.~Taniguchi}
    \affiliation{Advanced Materials Laboratory, National Institute for Materials Science, 1-1 Namiki, Tsukuba, 305-0044, Japan }
\author{J.~ Hone}
    \affiliation{Department of Mechanical Engineering, Columbia University, New York, New York 10027, USA}
\author{K.~L.~Shepard}
    \affiliation{Department of Electrical Engineering, Columbia University, New York, New York 10027, USA}
\author{P.~Kim}
    \affiliation{Department of Physics, Columbia University, New York, New York 10027, USA}
\date{\today}

\begin{abstract}
We report on a capacitance study of dual gated bilayer graphene.  The measured capacitance allows us to probe the electronic compressibility as a function of carrier density, temperature, and applied perpendicular electrical displacement $\DD$.  As a band gap is induced with increasing $\DD$, the compressibility minimum at charge neutrality becomes deeper but remains finite, suggesting the presence of localized states within the energy gap.  Temperature dependent capacitance measurements show that compressibility is sensitive to the intrinsic band gap. For large displacements, an additional peak appears in the compressibility as a function of density, corresponding to the presence of a 1-dimensional van Hove singularity (vHs) at the band edge arising from the quartic bilayer graphene band structure.  For $\DD>0$, the additional peak is observed only for electrons, while $\DD<0$ the peak appears only for holes. This asymmetry that can be understood in terms of the finite interlayer separation and may be useful as a direct probe of the layer polarization.
\end{abstract}
\maketitle

The unique band structures of monolayer (MLG) and bilayer graphene~\cite{Geim2007,CastroNeto2009} (BLG) offer unprecedented tunability in a high quality two dimensional electron system (2DES).  Within the independent electron approximation, both MLG and BLG are gapless, chiral systems. The gapless spectra are related to the pseudospin degeneracy, which is tied to the symmetry between the two sublattices constituting the honeycomb.  Whereas in the monolayer a gap can be opened only by a potential modulation on the spatial scale of the lattice  constant~\cite{Giovannetti2007}, in BLG the relevant sublattices are located on different layers, allowing a gap to be induced by a modulation of the interlayer imbalance \textit{via} the application of an electric field perpendicular to the BLG planes~\cite{McCann2006prl,McCann2006prb,Castro2007}. Although the field-effect tunable gap in BLG has been observed optically ~\cite{Ohta2006,Zhang2009a,Mak2009}, transport measurements show hopping conductivity at low temperatures.~\cite{Oostinga2008,Zou2010,Taychatanapat2010,Yan2010}

\begin{figure}[t]\includegraphics[width=1\linewidth,clip]{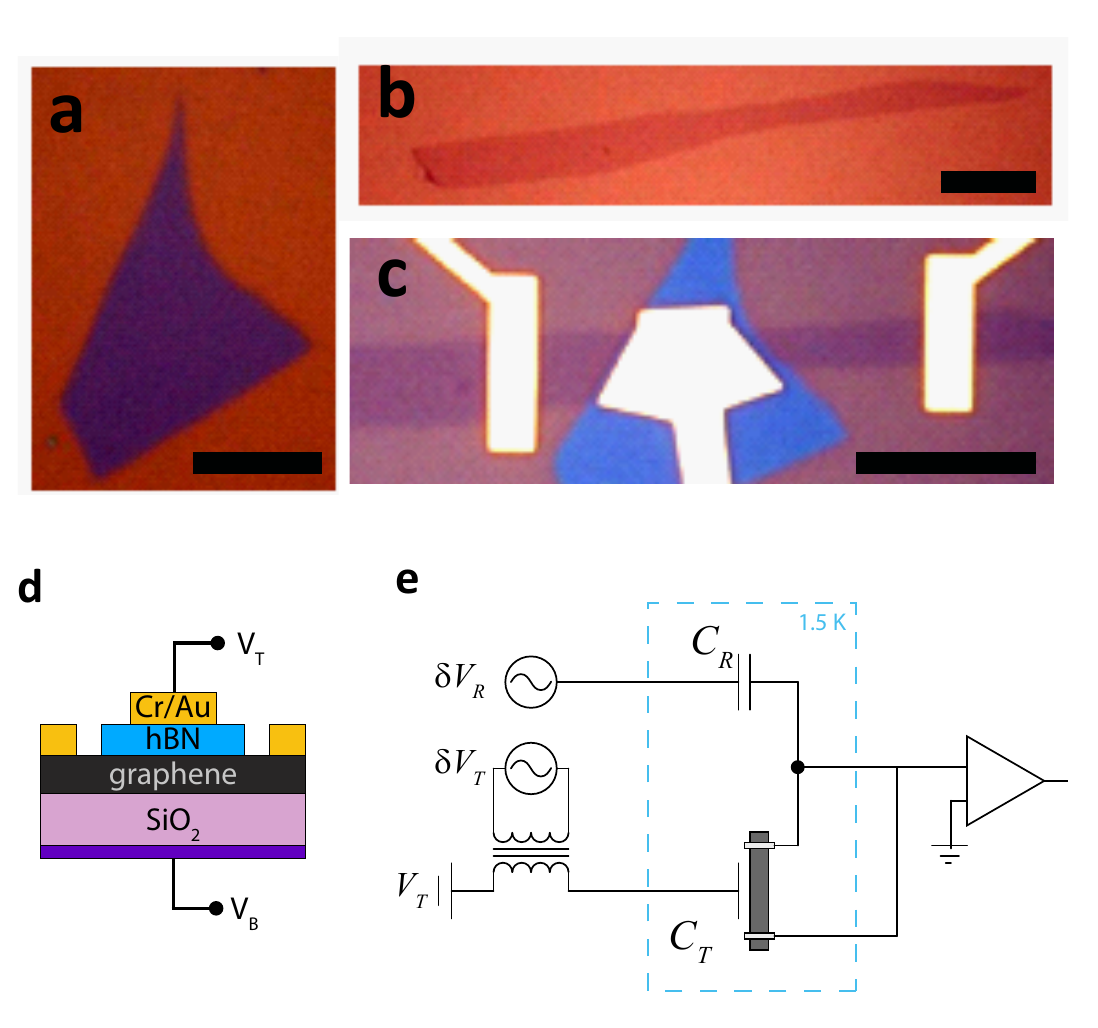}\caption{(a-c) Optical microscope images. (a) a thin hBN single crystal is transferred onto a (b) mono- or bilayer graphene flake and (c)  contacts and gate electrodes deposited by electron beam lithography. Scale bars are 10~$\mu$m.  (d) Cross section schematic of resulting dual gated device.  (e) Schematic circuit diagram of the capacitance bridge.  A reference capacitor (Johanson Technology R14S) is mounted on the probe, and a reference voltage is chosen to balance the capacitance bridge.  A small AC excitation signal is added to the DC gate bias through a transformer (Triad Magnetics SP67).
\label{f1}} \end{figure}

\begin{figure}[t]\includegraphics[width=\linewidth,clip]{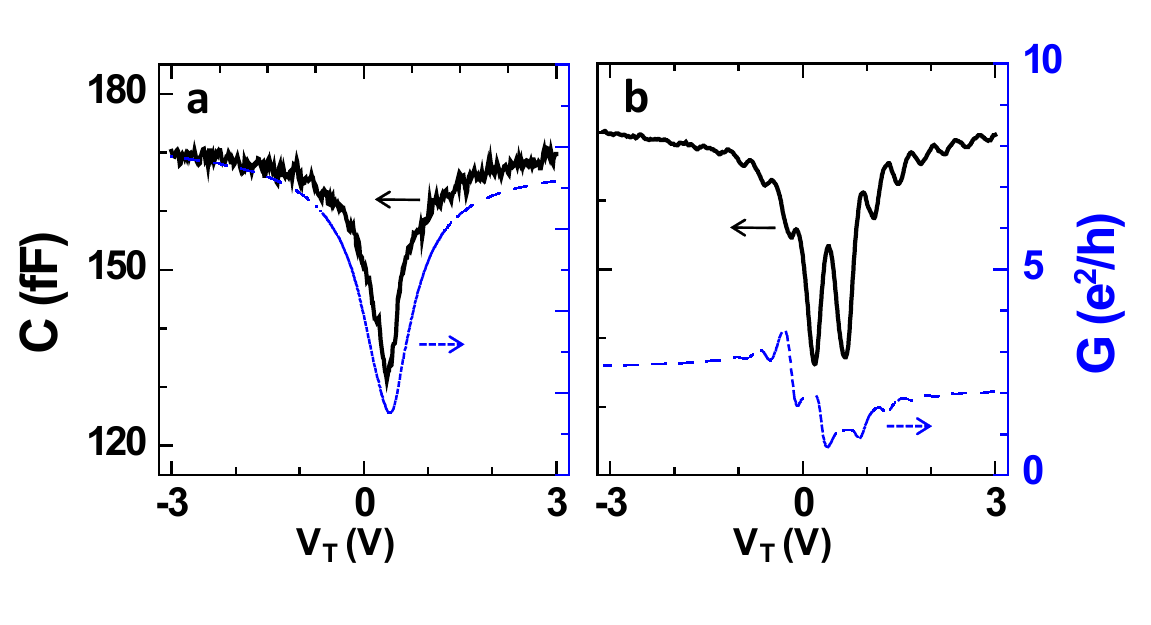}\caption{Measured conductance and capacitance of monolayer graphene at 2~K and $B=0$ (a) and $B=9$~T (b).  In (a), capacitance and conductance resemble each other closely due to the fact that both contain a spurious contribution adding ``in parallel''; $G=\left(1/R+1/R_C\right)^{-1}$ and $C_T=\left(1/C_Q+1/C_T^0\right)^{-1}$ where $R$ is the device resistance, $C_Q$ is the compressibility, or ``quantum capacitance,'' $C_T^0$ is the geometric capacitance, and $R_C$ is the contact resistance of our two-terminal devices.  In (b), capacitance reflects the electron \label{fmono}} \end{figure}

In a parallel plate capacitor made up of imperfect conductors, adding charge $n$ to the plates costs the sum of the classical electrostatic energy, the kinetic energy due to the resulting change in the chemical potential $\mu$, and the potential energy of Coulomb interactions between the charge carriers.  The measured differential capacitance, $C=\delta n/\delta V$, in such a system reflects this finite electronic compressibility by manifesting a lowered effective capacitance, $C^{-1}= C_0^{-1}+\left(e^2 A \nu\right)^{-1}$, where $C_0$ is the geometric capacitance, $A$ is the area of the device and $\nu \equiv \partial n/\partial \mu$ is the electronic compressibility, which corresponds to the density of states in the noninteracting, zero temperature limit.  In low dimensional systems the contribution of the compressibility to the capacitance---termed ``quantum capacitance''~\cite{Luryi1988}---can be small even when the conductivity remains large, providing a powerful tool in the study of both one \cite{Ilani2006} and two \cite{Eisenstein1992} dimensional electronic systems. Capacitance measurements are particularly powerful in the study of disordered systems, as they are able to detect localized states whose contribution to transport is suppressed.  Capacitance measurements are, as a result, crucial in understanding biased bilayer graphene, a system in which localization is known to play a role.  Moreover, the small but finite interlayer separation ($d\sim$~3.4~$\textrm{\AA}$) allows the layer polarization to be probed through electrostatic measurements, as described in a companion paper~\cite{Young2011}.

To produce dual gated graphene devices with high geometric capacitance, we utilize single crystal hexagonal boron nitride (h-BN) flakes~\cite{Kubota2007} as the top gate dielectric fabricated by the process described in Ref. \onlinecite{Dean2010}. Briefly, both graphene and single crystal h-BN, an insulating isomorph of graphite, are exfoliated onto n-Si/SiO$_2$ wafers (Fig. \ref{f1}(a-b)).
A thin (5-7 nm) h-BN flake is transferred on top of the graphene using a wet etch process and micromechanical manipulation~\cite{Jiao2009nl}, followed by electron beam lithography to form contact electrodes and a local top gate (Fig. \ref{f1}c).  The heavily doped silicon substrate, coated with 285 nm oxide, serves as the bottom gate. The double gated geometry allows independent control of the electronic density, $n$, and the displacement, $\DD=\frac{\varepsilon_B}{d_B}\left(\left(V_B-V_B^0\right)-\frac{C_T^0}{C_B^0}\left(V_T-V_T^0\right)\right)$ through the top and bottom gate voltages, $V_T$ and  $V_B$.  Here, $\varepsilon_B$ and $d_B$ are the dielectric constant and thickness of the back gate dielectric layer, $C_{T(B)}^0$ is the geometric capacitance of the top (bottom) gate, and $V_{T(B)}^0$ is the voltage offset required to obtain minimal density and displacement in the dual-gated region. We find that h-BN is an excellent gate dielectric, with $\varepsilon\sim$~3-4 and breakdown fields comparable ($\sim$.8 V/nm) to SiO$_2$ thin films.  In addition, we observe minimal degradation of graphene samples, with no additional doping contributed by the presence of the top gate and typical post h-BN transfer mobilities of $\mu\sim$5,000-10,000 cm$^2$/V~sec for graphene monolayers and $\mu\sim$2,000-3,000 cm$^2$/V sec for bilayers.

 Low temperature capacitance measurements were performed using a capacitance bridge circuit (Fig. 1(e)) with a cold reference capacitor.  All wires were shielded, and the sample package was encased in a Faraday cage to further reduce parasitic capacitances, which represent an additive constant to the measured value of $C_T$. A ceramic multilayer capacitor with minimal temperature dependence was chosen for the reference capacitor ($C_R$) and connected near the sample at low temperature. The noise level of the bridge was $\sim$25 $e/\sqrt{Hz}$, allowing sub-femtofarad resolution with averaging times of less than 30 seconds for our typical top gate AC excitation voltage $\delta V_T\sim$15-50~mV. Measurements were performed off-balance.  For the variable temperature capacitance measurements, we applied $\delta V_T=$50 mV AC excitation voltage on top of the DC gate bias and measured the current through the graphene device directly.  Although this method results in poorer signal to noise than the bridge measurement, it eliminates calibration errors stemming from the small temperature dependence of the reference capacitor.  The ability to measure the quantum contributions to the capacitance relies on the use of a thin top gate dielectric layer.  The 7 nm thick hBN dielectric results in values of $C_T^0$ that are comparable to the quantum capacitance $C_Q$, so that variations in the measured $C_T$ with changing electronic compressibility are easily detectable.

We check the efficacy of this measurement scheme using monolayer graphene capacitors.   MLG is expected to display monotonically increasing $\nu$ as a function of absolute density $|n|$.  In the presence of a strong magnetic field, $\nu$ is further modulated due to the formation of Landau levels.  Fig.~\ref{fmono} shows the measured capacitance $C$ and conductance $G$ of MLG as a function of top gate voltage $V_{T}$ at both zero and finite magnetic field.  The lowered compressibility stemming from the linear spectrum of of MLG can be inferred from a depression in the capacitance at zero density, while at
high $B$ the formation of the zero Landau level leads to a peak at charge neutrality~\cite{Martin2008}. The high magnetic field capacitance traces show compressibility oscillations due to the formation of higher LLs, while conductance shows the electron-hole asymmetry that is the signature of edge state transport in graphene heterojunctions~\cite{Abanin2007,Williams2007,Ozyilmaz2007prl}, which form due to the partial coverage of the graphene channel by the top gate.  The peak inversion at zero density and electron-hole symmetry in high magnetic field together confirm that the measured signal is indeed capacitive, and is sensitive to density of states rather than lateral resistance effects.

\begin{figure*}[t]\includegraphics[width=1\linewidth,clip]{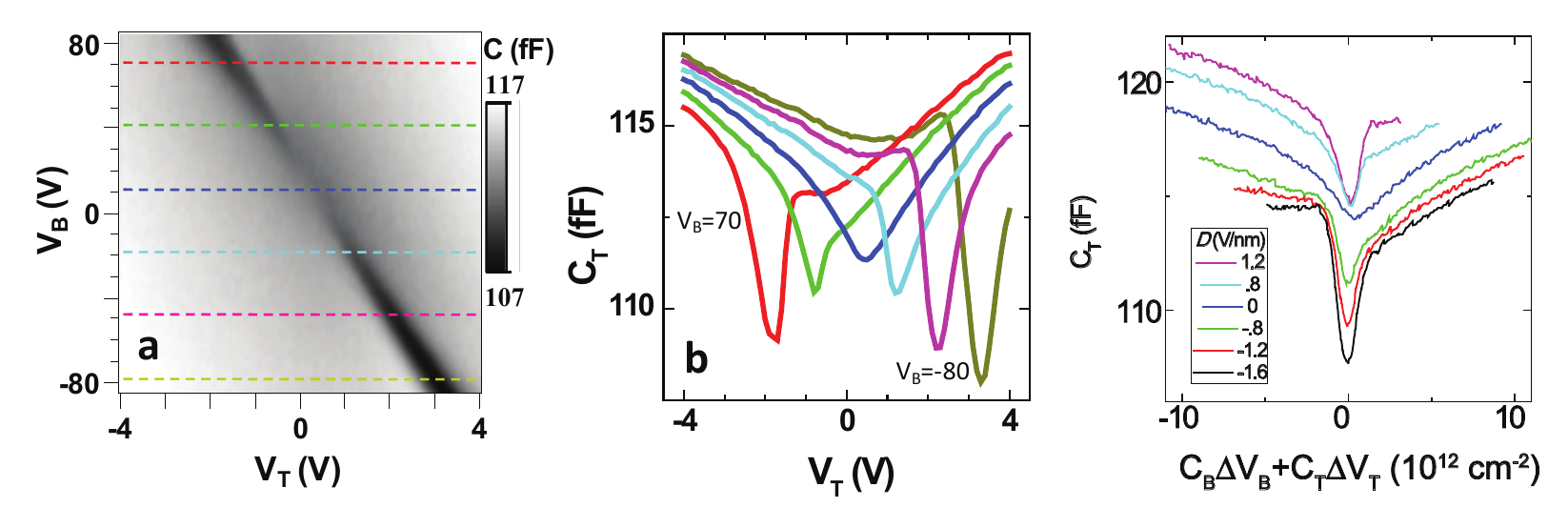}\caption{Capacitance at $B$=0 and 1.5 K as a function of V$_{T}$ and V$_{B}$.
Colored traces in (b) are taken at 30 volt intervals in V$_B$, corresponding to the colored lines in (a).  (c) Traces at constant $\overline{D}$, extracted from the data set shown in (a). Data is plotted as a function of $C_T^0 \Delta V_T+C_B^0 \Delta V_B$, which would correspond to the density were the bilayer perfectly 2 dimensional and perfectly compressible.  Curves in (c) are offset for clarity by 2fF per V/nm in $\DD$.
\label{f2}} \end{figure*}

We now turn our attention to bilayer graphene (BLG) samples. Fig. \ref{f2} shows the capacitance of a BLG sample at 1.5~K measured with the cold bridge.  Tuning external gates adjusts both $n$ and $\DD$.  For small values of $\DD\approx0$, the measured capacitance exhibits a minimum at $n=$~0 as expected for ungapped bilayer graphene, which has a hyperbolic band structure~\cite{Xia2009,Henriksen2010}. As $|\DD|$ increases, the $n=0$ minimum gets deeper, corresponding to the formation of a gap in the energy spectrum. The n=0 minimum does not go to zero at high values of $|\DD|$, and in fact the capacitance modulation is only 10\% with respect to the $\DD=0$ value.  In addition, a distinct local maximum develops next to the minimum.  As we argue below, the presence of both the dip and local peak in $\nu$ at high $|\DD|$ can be understood, at least qualitatively, from the band structure of gapped BLG, once the effects of disorder~\cite{Nilsson2007} and the interlayer separation~\cite{Young2011} are taken into account.

Within the nearest-neighbor tight binding approximation, the energy spectrum of pristine, Bernal stacked bilayer graphene with finite interlayer asymmetry $\Delta$ is gapped and has a ``Mexican hat'' structure~\cite{McCann2006prb,CastroNeto2009}.  Even in the presence of disorder, the absence of a positive quadratic term in the energy spectrum turns the problem of gapped, disordered bilayer graphene into one of a heavily doped semiconductor with \textit{quartic} energy bands~\cite{Mkhitaryan2008}.  We thus expect vestiges of a $\nu\propto 1/\sqrt{E}$ vHs-like feature to be present even in our low mobility bilayer samples, manifesting as a nonmonotonic-in-density feature at the band edge\cite{Nilsson2007,Mkhitaryan2008}.  In addition to smearing the band edge vHs, disorder has a dramatic effect on compressibility at charge neutrality in the presence of a large gap. In contrast to clean semiconductors, in which the depleted system is incompressible, our measured capacitance remains finite and large even for large $\overline{D}$, a fact we attribute to tails in the density of states representing localized intergap states~\cite{Mieghem1992}.  This explains the discrepancy between energy scales that govern transport~\cite{Oostinga2008,Zou2010,Taychatanapat2010,Yan2010} and the gap energies observed optically~\cite{Ohta2006,Zhang2009a,Mak2009}.  It also suggests that recently predicted topological edge conduction~\cite{Li2010} is not the dominant reason for incomplete ``turn-off'' in BLG devices of typical quality.

Quantitative analysis of the capacitance data requires extracting the compressibility $\nu$ from the measured signal, $C_T$.  For a perfectly two dimensional electron system ($C_{BL}\rightarrow\infty$), the measured top gate capacitance is
\begin{equation}
C_T^{-1}=\left(\frac{1}{C_T^0}+\frac{1}{A e^2\nu}\right)^{-1}+C_{para},
  \label{2Dcap}
\end{equation}
where $C_{para}$ are all parasitic capacitances between gate and contact electrodes and terms $\mathcal{O}\left(\frac{C_B^0}{C_T^0}\right)\sim.034$ have been neglected.  Extracting $\nu$ thus requires subtracting both parallel (C$_{para}$) and series (C$_T^0$) capacitances, and dividing by $A=$31$\mu$m$^2$ (determined by optical microscopy). We determined C$_T^0$/C$_B^0$ from the ratio of the back and top gate capacitances, measured by tracking the charge neutrality point in the V$_T$-V$_B$ plane (the dark diagonal belt in Fig. \ref{f2}(a)).  For the BLG device presented in this paper, we measured and $C_T/C_B=$29.5$\pm.4$, where $C_B=115$ ~aF$/\mu$m$^2$ is the geometric capacitance of the bottom gate.  As disordered BLG devices cannot be turned off completely, $C_{para}$ cannot be measured \textit{in situ} as is common practice in depletable semiconductors ~\cite{Eisenstein1990} and semiconducting carbon nanotubes~\cite{Ilani2006}.  Instead, we determine $C_P$=16$\pm$1~fF by removing the graphene through a short oxygen plasma etch and measuring the remaining capacitance between the metal contacts and the top gate.  $C_{para}$ constitutes about 10\% of the total capacitance signal.  Due to the subsequent subtraction of the (series) geometric capacitance $C_T^0$, the error in determination of $C_{para}$ is least important when the capacitance differs considerably from the geometric value.  This is the regime in which we perform a quantitative analysis of the compressibility of gapped BLG.

\begin{figure}[t]\includegraphics[width=1\linewidth,clip]{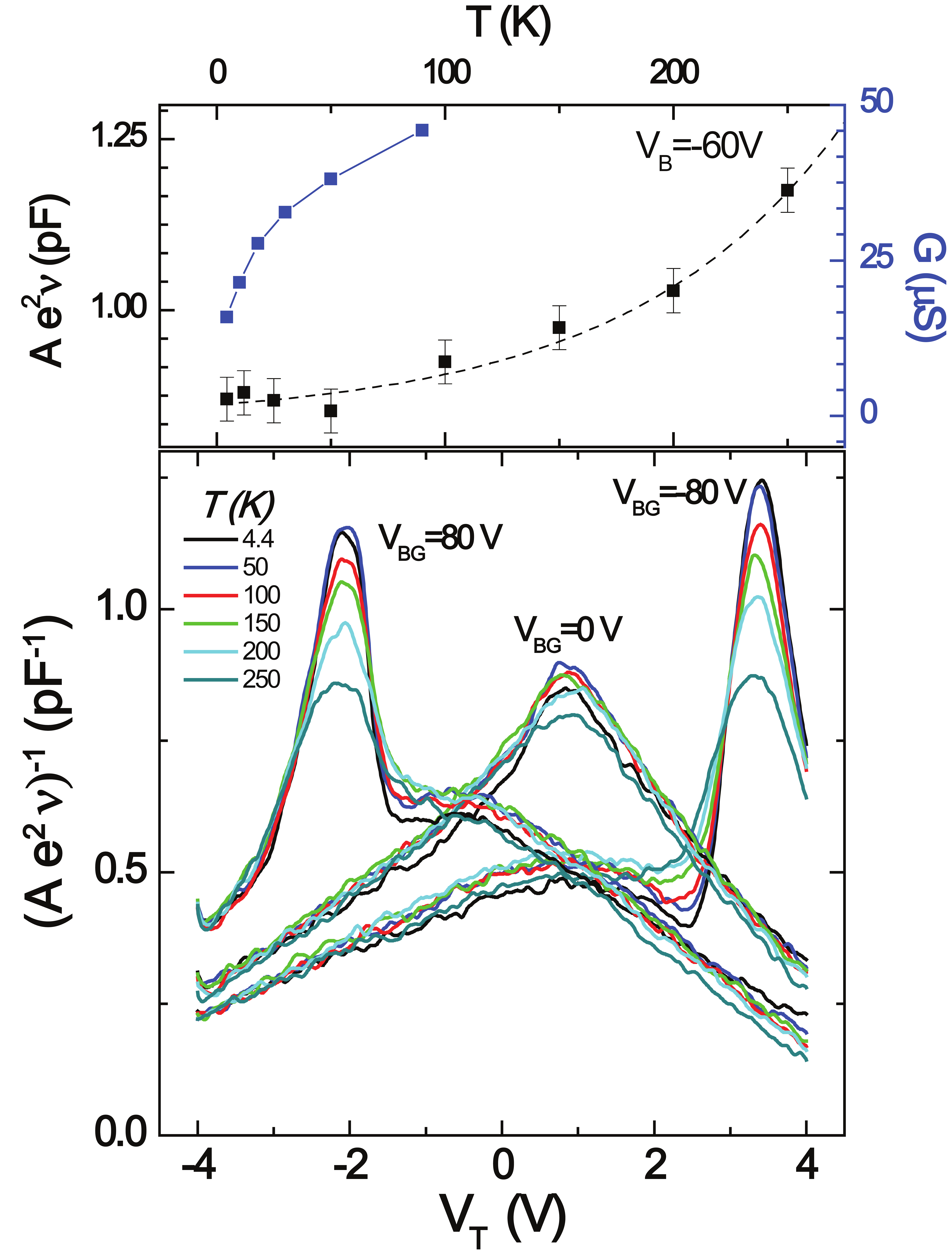}
\caption{Bottom panel: Temperature dependence of the inverse compressibility $(A e^2\nu)^{-1}$ for $V_{B}=$~80, 0 and -80 V. A single value of $C_{para}$ is chosen for all gate voltages at a given temperature, but a different $C_{para}$ is chosen for each temperature so that the curves match at high density.  We attribute the variation in $C_{para}$ to thermal expansion of the bonding wires, the capacitance of which constitute the majority of $C_{para}$.   Top panel: comparison between the temperature dependence of the minimal compressibility and minimal conductivity at V$_B$=-60 V. The dashed line is a guide for the eye.
\label{f4}} \end{figure}

Near overall charge neutrality at $|\overline{D}|\gg0$, our samples show a hopping conductivity similar to that observed in Ref. \onlinecite{Oostinga2008} from 1~K to $\sim$~150~K.  Capacitance instead shows no significant temperature dependence up to 50~K, thereafter slowly rising as might be expected for a spectrum with a gap in 50-100 meV range (Fig. \ref{f4}).  This is consistent with the presence of disorder-induced tails in the density of states throughout the band gap: whereas temperature dependent transport is dominated by the hopping between these localized states, temperature dependent capacitance is dominated by thermal population of the much larger density of states near the band edge.

The most interesting, and unexpected feature, of the experimental data are the local maxima observed at the band edge are associated with the 1D vHs inherent in the BLG band structure ~\cite{Nilsson2007,Mkhitaryan2008}. Interestingly, this feature is only present on one edge of the band, appearing on the electron side for $\overline{D}>0$ and the hole side for $\overline{D}<0$.  This inversion symmetry in the variables ($\overline{D}$,$n$) was observed in all devices measured, including those fabricated using a resist free shadow mask metallization as well as seeded atomic layer deposition of HfO$_2$~\cite{Farmer2009}.    Understanding this asymmetry requires taking into account the three dimensional structure of BLG~\cite{Young2011}, which consists of two strongly coupled but spatially distinct layers of carbon atoms.  The charge distribution on a BLG flake is sharply concentrated on the two layers, $n(z)\simeq n_1 \delta (z) +n_2 \delta(z+d)$, so that the system can be modeled as a four plate capacitor.  Solution of this electrostatic problem leads to the modified relation for the measured capacitance in which C$_T$ depends on both the \textit{interlayer} capacitance, $\nu_{21}$, and intralayer capacitances, $\nu_{11}$ and $\nu_{22}$, where $\nu_{ij}\equiv\partial n_i/\partial v_j$ with $n$ and $v$ represent layer indexed density and potential, and the indices i,j=1,2 denote the top or bottom layer. Crucially, while $\nu_{21}$ is symmetric with respect to layer interchange, $\frac{\partial n_2}{\partial v_2}$ is obviously asymmetric in the presence of interlayer asymmetry. Penetration field~\cite{Henriksen2010} measurements of bilayer graphene depend only on layer-symmetric quantities, and thus probe fundamentally different physical quantities.

\begin{figure}[t]\includegraphics[width=1\linewidth,clip]{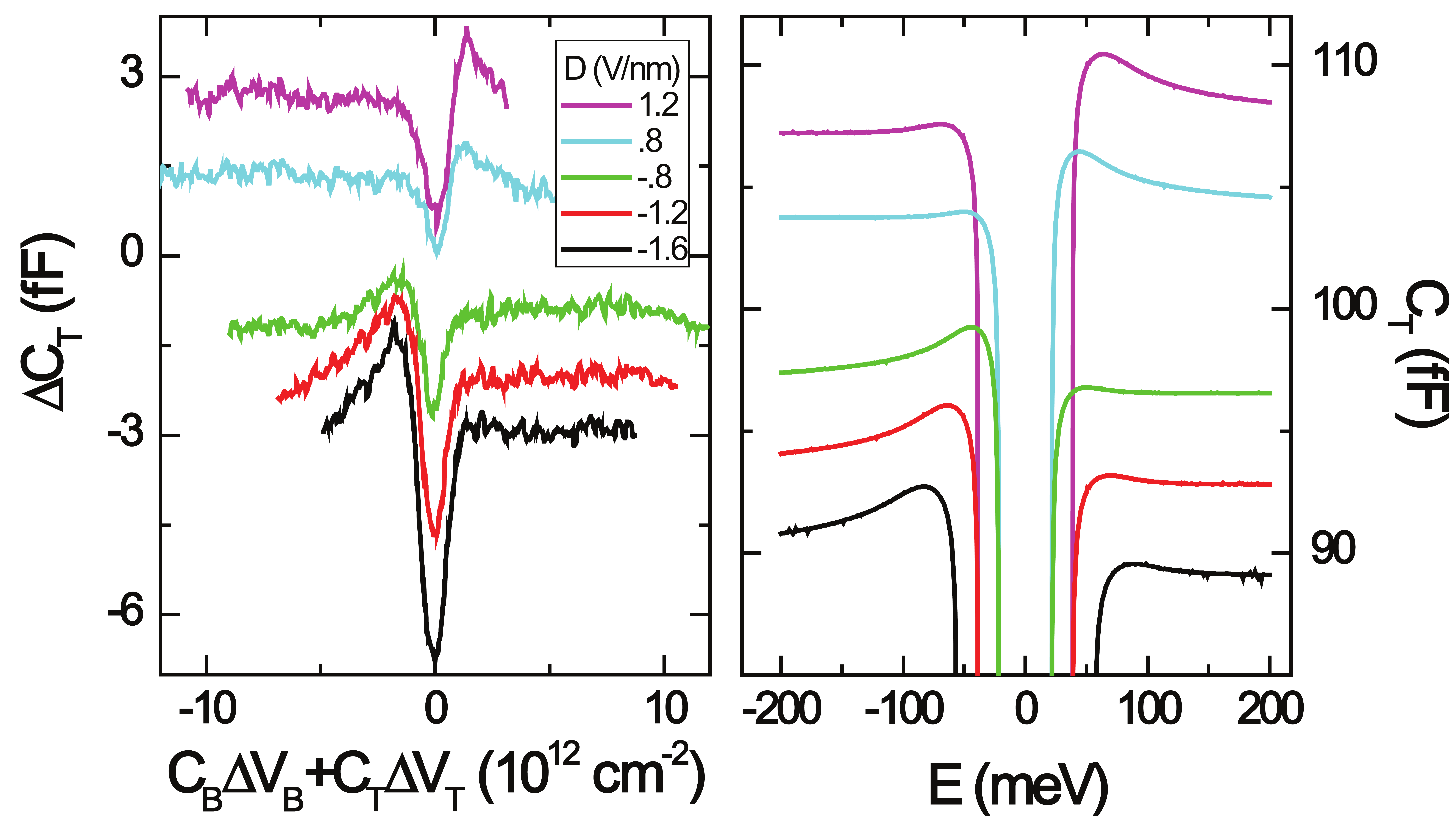}
\caption{Left Panel: Subtracted capacitance, $\Delta C_T=C_T(\DD)-C_T(\DD=0)$, as a function of approximate density for different applied displacements.  Curves are offset by 1fF per V/nm in $\DD$.  Right Panel: Calculated top gate capacitance for disordered bilayer graphene, following Ref. ~\onlinecite{Young2011}.  The colors correspond to the displacements in the left panel; the disorder parameter is $\gamma=4$ for all curves. Curves are offset by 3 fF per V/nm in $\DD$.  \label{f3}} \end{figure}

As elaborated in Ref. \onlinecite{Young2011}, the $1/\sqrt{E}$ divergence associated with the 1D vHs at the band edge principally on the low energy layer within the BLG flake.  The vHs manifests more strongly in the measured capacitance when vHs-hosting layer is closest to the top gate; conversely, the gate sees a far-layer vHs only through the screened field penetrating the near layer.  Counterintuitively, disorder enhances this effect not only by smearing the total density of states~\cite{Nilsson2007,Mkhitaryan2008} but by populating the normally depleted non-vHs layer, thereby enhancing its ability to screen.  While an ideal experimental geometry would permit the simultaneous measurement of capacitance from two sides of the BLG flake this is effectively accomplished in our single local gate geometry by reversing the sign of $\overline{D}$, thus reversing the order of the vHs bearing and non-bearing layers.

In order to better compare our experimental data with theory based on a parabolic two band model, it is convenient to subtract a background taken at $\overline {D}=0$. Because the high energy behavior depends only weakly on the displacement, this has the effect of isolating the low energy part of the measured capacitance, and in addition removes the effects of electron-hole asymmetric elements of the band structure\cite{Henriksen2010}.  The results of this subtraction resemble theoretical calculations which take into account both the interlayer separation as well as weak, short-range disorder (Fig. \ref{f3}). In particular, the asymmetric appearance of the van Hove singularity can be understood as the effect of disorder enhanced interlayer screening.  However, quantitative understanding of the role of disorder in bilayer graphene will require experiments that independently control the disorder, as well as a more sophisticated theory taking into account a wider variety of effects including long range scattering and electron-electron interactions.

In conclusion, we study the broken symmetry state of bilayer graphene induced by an electrical displacement field applied perpendicular to the bilayer.  We observed the formation of the displacement induced gap and its accompanying 1D vHs, and estimate the gap size from temperature dependent measurements.  We also discussed how different measurement geometries make capacitance a probe of layer-pseudospin polarization in the bilayer. In BLG, layer symmetry breaking in BLG can also be expected to occur spontaneously, as a result of electronic interactions at both finite~\cite{Barlas2008,Feldman2009,Zhao2010,Dean2010,Min2008prb2} and zero~\cite{Nandkishore2009} magnetic fields. Similar measurement performed on high-quality bilayers manifesting these effects will allow spontaneous layer polarization to be probed directly through comparison of measurements in various capacitance geometries, as discussed at length in a companion paper~\cite{Young2011}.

%%%%%%%%%%%%%%%%%%%%%%%%%%%%%%%%%%%%%%%%%%%%%%%%%%%%%%%%%%%%%%
\begin{acknowledgments}The authors acknowledge discussions with I.L. Aleiner, B.L. Altshuler, J.P. Eisenstein, E.A. Henriksen, L.S. Levitov, A.H. MacDonald, K.F. Mak, and E. McCann. This work is supported by AFOSR MURI and INDEX (A.Y.F and P. K.)).\end{acknowledgments}
%\bibliography{references}

\begin{thebibliography}{10}

\bibitem{Geim2007}
A.~K. Geim and K.~S. Novoselov, Nature Materials {\bf 6},  183  (2007).

\bibitem{CastroNeto2009}
A.~H. {Castro Neto} {\it et~al.}, Reviews of Modern Physics {\bf 81},  109
  (2009).

\bibitem{Giovannetti2007}
G. Giovannetti {\it et~al.}, Phys. Rev. B {\bf 76},  073103  (2007).

\bibitem{McCann2006prl}
E. McCann and V.~I. Fal'ko, Phys. Rev. Lett. {\bf 96},  086805  (2006).

\bibitem{McCann2006prb}
E. McCann, Phys. Rev. B {\bf 74,},  161403  (2006).

\bibitem{Castro2007}
E.~V. Castro {\it et~al.}, Phys. Rev. Lett. {\bf 99},  216802  (2007).

\bibitem{Ohta2006}
T. Ohta {\it et~al.}, Science {\bf 313},  951  (2006).

\bibitem{Zhang2009a}
Y. Zhang {\it et~al.}, Nature {\bf 459},  820  (2009).

\bibitem{Mak2009}
K.~F. Mak, C.~H. Lui, J. Shan, and T.~F. Heinz, Phys. Rev. Lett. {\bf 102},
  256405  (2009).

\bibitem{Oostinga2008}
J.~B. Oostinga {\it et~al.}, Nature Materials {\bf 7},  151  (2008).

\bibitem{Zou2010}
K. Zou and J. Zhu, Phys. Rev. B {\bf 82},  081407  (2010).

\bibitem{Taychatanapat2010}
T. Taychatanapat and P. Jarillo-Herrero, Phys. Rev. Lett. {\bf 105},  166601
  (2010).

\bibitem{Yan2010}
J. Yan and M.~S. Fuhrer, Nano Lett. {\bf 10},  4521  (2010).

\bibitem{Luryi1988}
S. Luryi, Applied Physics Letters {\bf 52},  501  (1988).

\bibitem{Ilani2006}
S. Ilani, L.~A.~K. Donev, M. Kindermann, and P.~L. McEuen, Nature Physics {\bf
  2},  687  (2006).

\bibitem{Eisenstein1992}
J.~P. Eisenstein, L.~N. Pfeiffer, and K.~W. West, Phys. Rev. Lett. {\bf 68},
  674  (1992).

\bibitem{Young2011}
A.~F. Young and L.~S. Levitov, Phys. Rev. B {\bf 84},  085441  (2011).

\bibitem{Kubota2007}
Y. Kubota, K. Watanabe, O. Tsuda, and T. Taniguchi, Science {\bf 317},  932
  (2007).

\bibitem{Dean2010}
C.~R. Dean {\it et~al.}, Nature Nanotechnology {\bf 5},  722  (2010).

\bibitem{Jiao2009nl}
L. Jiao {\it et~al.}, Nano Letters {\bf 9},  205  (2009).

\bibitem{Martin2008}
J. Martin {\it et~al.}, Nature Physics {\bf 4},  144  (2008).

\bibitem{Abanin2007}
D.~A. Abanin and L.~S. Levitov, Science {\bf 317},  641  (2007).

\bibitem{Williams2007}
J.~R. Williams, L. DiCarlo, and C.~M. Marcus, Science {\bf 317},  638  (2007).

\bibitem{Ozyilmaz2007prl}
B. \"{O}zyilmaz {\it et~al.}, Phys. Rev. Lett. {\bf 99},  166804  (2007).

\bibitem{Xia2009}
J. Xia, F. Chen, J. Li, and N. Tao, Nature Nanotechnology {\bf 4},  505
  (2009).

\bibitem{Henriksen2010}
E.~A. Henriksen and J.~P. Eisenstein, Phys. Rev. B {\bf 82},  041412  (2010).

\bibitem{Nilsson2007}
J. Nilsson and A.~H. {Castro Neto}, Phys. Rev. Lett. {\bf 98},  126801  (2007).

\bibitem{Mkhitaryan2008}
V.~V. Mkhitaryan and M.~E. Raikh, Phys. Rev. B {\bf 78},  195409  (2008).

\bibitem{Mieghem1992}
P. Van~Mieghem, Rev. Mod. Phys. {\bf 64},  755  (1992).

\bibitem{Li2010}
J. Li, I. Martin, M. Buttiker, and A.~F. Morpurgo, Nat Phys {\bf 7},  38
  (2011).

\bibitem{Eisenstein1990}
J.~P. Eisenstein, H.~L. Stormer, L.~N. Pfeiffer, and K.~W. West, Phys. Rev. B
  {\bf 41},  7910  (1990).

\bibitem{Farmer2009}
D.~B. Farmer {\it et~al.}, Nano Letters {\bf 9},  4474  (2009).

\bibitem{Barlas2008}
Y. Barlas, R. C\^ot\'e, K. Nomura, and A.~H. MacDonald, Phys. Rev. Lett. {\bf
  101},  097601  (2008).

\bibitem{Feldman2009}
B.~E. Feldman, J. Martin, and A. Yacoby, Nature Physics {\bf 5},  889  (2009).

\bibitem{Zhao2010}
Y. Zhao, P. Cadden-Zimansky, Z. Jiang, and P. Kim, Phys. Rev. Lett. {\bf 104},
  066801  (2010).

\bibitem{Min2008prb2}
H. Min, G. Borghi, M. Polini, and A.~H. MacDonald, Phys. Rev. B {\bf 77},
  041407  (2008).

\bibitem{Nandkishore2009}
R. Nandkishore and L. Levitov, Phys. Rev. Lett. {\bf 104},  156803  (2010).

\end{thebibliography}
%\bibliographystyle{prsty}

\end{document}